\documentstyle[12pt]{article}
\newcommand{\scite}{~\cite}

\def\beq{\begin{equation}}
\def\eeq{\end{equation}}
\def\beqa{\begin{eqnarray}}
\def\eeqa{\end{eqnarray}}
\def\rd{{\mathrm d}}

\def\as{\alpha_s}

\def\msbar{{\overline{MS}}}
\newcommand\aem{\alpha_{\rm em}}
\newcommand\sss{\scriptscriptstyle}

\newcommand\mug{\mu_\gamma}

\newcommand\muf{\mu_{\sss F}}
\newcommand\mur{\mu_{\sss R}}

\def\pt{p_T}

\begin{document}

\begin{titlepage}

\begin{flushright}
\begin{tabular}{l}
             DESY 95-103\\
             FNT/T-95/12 \\
             LNF-95/025(P)\\
             hep-ph/9505419 \\
             May 1995\\
\end{tabular}
\end{flushright}
\vspace{2.5cm}

\begin{center}
{\huge Charm Photoproduction \\via Fragmentation} \\
\vspace{2.5cm}
{\large Matteo Cacciari$^a$\footnote{Della Riccia fellow. Work supported by
INFN - Pavia (Italy)}
and Mario Greco$^b$
} \\
\vspace{1.cm}
{\sl $^a$DESY - Deutsches Elektronen-Synchrotron, Hamburg, Germany\\
E-mail: cacciari@desy.de \\
\vspace{.3cm}
$^b$Dipartimento di Fisica,
Universit\`a di Roma III, and \\
INFN, Laboratori Nazionali di Frascati, Italy \\
E-mail: greco@lnf.infn.it }
\vspace{2.5cm}
\begin{abstract}
The next-to-leading open charm production in $\gamma p$
collisions is
calculated within the Perturbative Fragmentation Functions formalism, to allow
resummation of $\as\log(\pt^2/m^2)$ terms. In the
large $\pt$ region $(\pt > m)$ the result is consistent with the
fixed order NLO calculation, small discrepancies being found for very large
$\pt$ and at the edge of phase space. The two approaches differ in the
definition and the relative
contribution of the direct and resolved terms, but essentially agree on their
sum. The resummation is found to lead to a
reduced sensitivity to the choice of the
renormalization/factorization scale.
\end{abstract}

\end{center}

\end{titlepage}

\section{Introduction}

The inclusive photoproduction of heavy quarks, namely the process
\begin{eqnarray*}
\gamma + H \to Q + X,
\end{eqnarray*}
has been evaluated up to next-to-leading order (NLO) in recent years.
R.K. Ellis and
P. Nason have first evaluated the QCD corrections to the direct
$\gamma H\to QX$ process\scite{EN}. Their calculation has been successively
confirmed by J.~Smith
and W.L.~van Neerven\scite{smith}. The NLO corrections for parton-parton
scattering,
needed for the resolved part, have  been produced by P. Nason, S. Dawson
and R.K. Ellis\scite{NDE} and also by W. Beenakker {\it et al.}\scite{wim}

More recently S. Frixione  et al. have
finally written computer codes\scite{mnr,fmnr}
capable of integrating the
NLO formulas thereby producing phenomenologically useful cross sections.
These codes will be used to perform comparisons with our results.

Within the context of NLO calculations of heavy quark photoproduction, it is
also worth reminding the calculation of inelastic $J\!/\!\psi$ photoproduction
which has recently been presented by M. Kr\"amer {\it et al.}\scite{michael}.

All the above fixed-order perturbative calculations rely on the fact that
the mass of the heavy quark acts as a cutoff
for infrared singularities. The splitting
processes which involve a heavy quark are therefore finite order by order in
perturbation theory: there is no need to subtract singularities when
handling a heavy quark line. This has induced the authors of refs.\cite{EN,NDE}
to adopt a modified version\scite{cwz} of the $\msbar$ subtraction scheme
in which the
heavy quark effects decouple in processes involving momenta much smaller than
the heavy quark mass. This also implies that the heavy quark does not
contribute
to the evolution of the running coupling and of the structure functions, and
therefore the absence of subprocesses initiated by an intrinsic heavy
flavour generated by the structure function. All the effects of the
heavy quark are therefore contained in the kernel cross section only.

The NLO one particle inclusive differential distribution will however contain
terms
of the kind $\as\log(\pt^2/m^2)$ which, in the large $\pt$ limit, will become
large and will spoil the perturbative expansion of the cross section. This is
reflected in a large sensitivity to the choice of the
renormalization/factorization (r/f) scales and hence in a large uncertainty
in the
theoretical prediction. Similar potentially large logarithms, of the kind
$\as\log(Q/m)$, $Q$ being the photon virtuality, do appear in NLO
calculations of heavy quark electroproduction total cross sections. The
resummation of these logs is being considered in a series of
papers\scite{olness}.

In Ref.~\cite{cg} the problem of large $\as\log(\pt^2/m^2)$ terms was
tackled in the hadroproduction case
by introducing the technique of the Perturbative Fragmentation Functions (PFF)
through which these terms were resummed to all orders and the cross section was
shown to display a milder scale sensitivity.

More in detail, the PFF approach is based on the assumption that when the
momenta involved are much larger than
their mass, the heavy flavours behave as if they were
massless. Technically, mass terms in the kernel cross sections are suppressed
by
powers of $m/\pt$. This means that one can go back to the usual $\msbar$ scheme
and treat all the quarks  as massless, thereby subtracting also the
singularities of the heavy quark lines and introducing a fragmentation
function to absorb the final state divergencies. The key point is that the
massiveness of the heavy quark can be exploited to calculate in perturbative
QCD
the initial state conditions for the heavy quark fragmentation functions at a
scale of the order of its mass\scite{melenason}. Therefore
the PFF approach relies on the same
input parameters of the perturbative calculation, but allows a resummation
of large logarithms of $(\pt/m)$. However
the neglecting of the mass terms will of course make the PFF result not
accurate
in the region $\pt\sim m$. In particular, the PFF approach is not able to
calculate the total cross section, which is obtained from the fixed-order
calculation.

\section{Scales in the photoproduction process}

As well known, when calculating the next-to-leading order correction to a
direct
photoproduction process (i.e. the ${\cal O}(\aem\as^2)$ terms) one finds that
singular terms associated with the electromagnetic vertex do appear.
They are due to the emission of collinear light (massless) partons from the
incoming photon and they are a signal for the presence of a
non-perturbative region where the photon splits into quarks and
gluons before interacting with the partons in the hadronic
target. This problem is solved in very much the same way as the hadroproduction
case: the collinear singularity is subtracted at a given scale and factored
into
a structure function. The photoproduction process gets therefore splitted into
two pieces.

Above the aforementioned scale the photon couples with
pointlike coupling
to a quark which then undergoes a hard
scattering. This term is called the point-like or direct
contribution.
Below the factorization scale the photon is regarded as
a hadron, and its structure function gives the probability for it emitting a
parton which subsequently participates in the hard scattering process. Hence
the
name of hadronic or resolved photon component.

The photon structure functions will of course depend upon the momentum scale
$\mu_\gamma$ at which the collinear singularities of the photon leg
are subtracted.
Neither the point-like nor the hadronic components are separately
independent of $\mu_\gamma$, because the subtracted term
in the direct component is related to the definition
of the photon structure function in the resolved component.

Let us now consider the generic expression for the heavy-quark production
at large $\pt$ via fragmentation due to
an on-shell photon colliding with a hadron $H$.
Expliciting the various scale dependencies in the
process, we write the ${\cal O}(\aem\as^2)$ cross section in the following form
\beqa
&&\rd\sigma_{\gamma {\sss H}\to QX}\sim
\sum_{ik}\int {\rd x\,\rd z}\,F_H^i(x,\muf)\rd\hat{\sigma}_{\gamma i\to k}
  (x,z,\as(\mur),\mur,\muf,\mug) D_k^Q(z,\muf)\nonumber \\
&&\quad
+ \sum_{ijk}\int \rd x_1\,\rd x_2\,\rd z\, F_{\gamma}^i(x_1,\muf^\prime,\mug)
F_H^j(x_2,\muf^\prime) \rd\hat{\sigma}_{ij\to k}
  (x_1,x_2,z,\as(\mur^\prime),\mur^\prime,\muf^\prime,\mug)D_k^Q(z,\muf^\prime)
\nonumber\\
&&
\label{sigmascales}
\eeqa
with, at the NLO here considered,
\beqa
\rd\hat{\sigma}_{\gamma i\to k}
  (\as(\mur),\mur,\muf,\mug)&=&
   \aem\as(\mur)\rd\hat{\sigma}^{(0)}_{\gamma i\to k}
  +\aem\as^2(\mur)\rd\hat{\sigma}^{(1)}_{\gamma i\to k}
  (\mur,\muf,\mug) \nonumber \\
\rd\hat{\sigma}_{ij\to k}
  (\as(\mur),\mur,\muf,\mug)&=&
   \as^2(\mur)\rd\hat{\sigma}^{(0)}_{ij\to k}
  + \as^3(\mur)\rd\hat{\sigma}^{(1)}_{ij\to k}
  (\mur,\muf,\mug)
\eeqa
Here $\mur$ and $\mur^\prime$ are renormalization scales,
$\muf$ and $\muf^\prime$ are factorization scales for collinear
singularities arising from strong interactions,
and $\mug$ is a
factorization scale for collinear singularities arising from
the electromagnetic vertex. All the structure and fragmentation functions
obey the usual Altarelli-Parisi evolution equations
 \beq
{{\rd F^i(x,\muf)}\over{\rd\log\muf^2}} = {{\as(\muf)}\over{2\pi}}\int_x^1
{{\rd y}\over y} P_{ki}(x) F^k(y/x,\muf)
\label{glap}
\eeq
with the exception of the photon structure function $F_\gamma^i$ which also
has an inhomogeneous evolution in $\mug$:
\beq \label{inhomogeneous}
\frac{\partial F_\gamma^i(x,\muf,\mug)}{\partial\log\mug^2}=
\frac{\aem}{2\pi} e_i^2
\left[x^2+(1-x)^2\right] + {\cal O}(\as),
\eeq
In the commonly-used photon density parametrizations,
$\mug$ is usually kept equal to $\muf$, so that the term given in
eq.~(\ref{inhomogeneous})
becomes a correction to the usual Altarelli-Parisi equation (the
so-called inhomogeneous term).

The cross section (\ref{sigmascales}) is of course independent of the
renormalization/factorization scales at the perturbative order at which it is
calculated. Indeed $\mur$ cancels between $\as(\mur)$
and the explicit dependence of the
$\sigma^{(1)}$ kernels; $\muf$ cancels between the structure/fragmentation
functions and again the NLO kernel dependencies; finally, $\mug$ cancels
between
the pointlike component and the photon structure function evolution in the
resolved component. This last cancellation, in particular, prompts for the need
to consider both the pointlike and the resolved component when evaluating
photoproduction cross section. Being the factorization procedure entirely
arbitrary, none of the two components has any physical meaning, only their sum
being observable\footnote{For a detailed discussion on the cancellation of
the $\mug$ dependence in the photoproduction of jets, see ref.\cite{gv}}.

\section{The PFF approach}

We will consider now in detail the photoproduction of heavy quarks in the
framework of PFF. Then, following for instance ref.\cite{ellis}, we can rewrite
the multidifferential cross
sections for the photoproduction process
\begin{eqnarray*}
\gamma(P_1) + H(P_2) \to Q(P) + X
\end{eqnarray*}
as follows. The resolved part reads
\beqa
{{E\rd^3\sigma^{res}}\over{\rd^3P}} &=& {1\over{\pi S}}
\sum_{ijk}\int^1_{1-V+VW}
{{\rd z}\over{z^2}}
\int^{1-(1-V)/z}_{VW/z} {{\rd v}\over{1-v}} \int^1_{VW/zv} {{\rd w}\over w}
\times\nonumber\\
&&\times F_H^i(x_1,\muf)F_H^j(x_2,\muf)
D_k^Q(z,\muf)\times\label{resolved}
\\
&&\times \left[{1\over v}\left({{\rd\sigma^0(s,v)}\over{\rd v}}\right)_{ij\to
k}
\delta(1-w) + {{\as^3(\mur)}\over{2\pi}}K_{ij\to
k}(s,v,w;\mur,\muf)\right]\nonumber
\eeqa
having defined the hadron-level quantities
\beq
V = 1+{T\over S}\qquad\qquad W = {-U\over{S+T}}
\eeq
with $S=(P_1+P_2)^2$, $T=(P_1-P)^2$ e $U=(P_2-P)^2$. Similarly are defined the
parton-level ones $s$, $v$ e $w$. In terms of the momentum fractions $x_1$,
$x_2$ and $z$ it holds
\beq
s = x_1 x_2 S\qquad x_1 = {{VW}\over{zvw}}\qquad x_2 = {{1-V}\over{z(1-v)}}
\eeq
Notice that we have abandoned any distinction between the photon and the
hadron factorization
scales $\mug$ and $\muf$. In the following we'll also usually
identify the renormalization and the factorization scales, i.e. $\mur = \muf =
\mu$.

The expression for the pointlike contribution amounts to rewriting
(\ref{resolved}) with the constraint $F_\gamma^i(x,\muf) = \delta(1-x)$ with
kernel cross sections for $\gamma$ scattering.
By exchanging the integrations over $v$ and $w$ we can write
\beqa
{{E\rd^3\sigma^{point}}\over{\rd^3P}} &=& {1\over{\pi S}}
\sum_{ik}\int^1_{1-V+VW}
{{\rd z}\over{z^2}}
\int^1_{VW/(z-1+V)} {{\rd w}\over w} \int {{\rd v}\over{1-v}}
F_H^i(x,\muf)
D_k^Q(z,\muf)\times
\nonumber\\
&&\times \left[{1\over v}\left({{\rd\sigma^0(s,v)}\over{\rd v}}\right)_{\gamma
i\to k}
{{VW}\over{zw}}\delta\left(v - {{VW}\over{zw}}\right)\delta(1-w)+
\right.\label{pointlike}\\
&&+\left.
{{\as^2(\mur)}\over{2\pi}}K_{\gamma i\to
k}(s,v,w;\mur,\muf){{VW}\over{zw}}\delta\left(v - {{VW}\over{zw}}\right)
\right]\nonumber
\eeqa
with
\beq
 s = xS \qquad\qquad x = {{1-V}\over{z-VW/w}},
\eeq
and we have dropped the integration limits for $v$, being this integration
constrained by the $\delta$-function.

We recall that the PFF approach to large-$\pt$ heavy quark production assumes
that the heavy quark is produced via fragmentation of partons (charm included)
which have been
produced in a massless way in the hard scattering. The fragmentation functions
$D_k^Q$ can be evaluated in perturbative QCD at a scale of the order of the
heavy quark mass and subsequently evolved up to the desired factorization scale
$\muf$ through the Altarelli-Parisi equations.

The lowest order massless kernel cross sections $\rd\sigma^0_{ij\to k}$ and
$\rd\sigma^0_{\gamma i\to k}$ (of order $\as^2$ and $\aem\as$ respectively)
have been long known in literature. The next-to-leading terms for
massless partons scattering
$K_{ij\to k}$ (of order $\as^3$) and $K_{\gamma i\to k}$
(of order $\aem\as^2$) have been instead produced in recent years.
Indeed
Aversa et al.\scite{aversa} have given the explicit
expressions for the NLO massless parton-parton scattering processes,
$K_{ij\to k}$,
while
Aurenche et al.\scite{aurenche}
have instead calculated the NLO
corrections to the $\gamma$-massless parton scattering process,
$K_{\gamma i\to k}$.

For the numerical evaluation of eqs. (5) and (8)
we have used the computer programs originally developed by those
authors\scite{aversa,aurenche} to perform the convolution of the scattering
kernels with the structure
and fragmentation functions. Some modifications had however to be implemented
to ensure
the numerical convergence of the integrals. The codes had actually been
designed
to handle light hadrons fragmentation functions. These functions typically tend
rapidly to zero for increasing values of the argument $z$, and can therefore
easily be numerically integrated. This is no more true when treating heavy
quarks due to the singular behaviour of the $D_Q^Q$ fragmentation function
in $z=1$. In
ref.~\cite{cg} a so called ``pole subtraction'' procedure was implemented. In
this case we have instead modified the fragmentation function by convoluting it
with a
``regulator function'' $D_{rf}(x;\alpha,\beta)$ given by
\beq
D_{rf}(x;\alpha,\beta) = A\; (1-x)^\alpha x^\beta
\eeq
$\alpha$ and $\beta$ have to be chosen such that $D_{rf}(x)$ is peaked very
near $x=1$ (we have taken $\alpha = 1$ and $\beta = 1000$),
and $A$ is the normalization factor:
\beq
A^{-1} = \int_0^1\rd x\;D_{rf}(x;\alpha,\beta) = B(\beta,\alpha+1)
\eeq
$B$ being the Euler beta-function.
The $D_{rf}(x;\alpha,\beta)$ goes to zero in $x=1$ fast enough
to allow the
numerical integration.
We have explicitly checked that the numerical result is
compatible - within errors - with the ``exact'' one given by the
pole subtraction method.

A further modification of the codes concerns the effect of
the heavy quark mass $m$ on the renormalization/factorization
scales and on the kinematical boundaries of the integrals.
Indeed the heavy quark mass has
been taken explicitly into account by using
\beq
\mu_{ref} = \sqrt{\pt^2 + m^2}
\label{muref}
\eeq
as the central choice for the f/r scale and also
by calculating the $S$, $T$ and $U$
invariant with massive kinematic. Given the heavy quark transverse momentum
$\pt$ and its scattering angle in the center of mass frame $\theta$ this
amounts to write for the heavy
quark quadrimomentum $P$
\beq
P = \left(\sqrt{{\pt^2\over{\sin^2\theta}} + m^2}; \pt,0,
{{\pt\cos\theta}\over{\sin\theta}}\right)
\eeq
These simple kinematical considerations have of course some clear effect
in the low-$\pt$ region, making the numerical integration easier,
but in no way allow one to reproduce the result of the full massive
calculation,
being the O($m/\pt$) terms in the kernel cross sections still missing.

\section{Numerical results and conclusions}

Let us consider the inclusive photoproduction of a $c$ quark at
present HERA energy.

For the sake of simplicity we'll take a photon of fixed energy
$E_\gamma =$ 26.7 GeV which scatters against a
$E_p =$ 820 GeV proton. This amounts to a center of mass energy of 296 GeV.
According to the usual conventions, we'll take positive rapidities in the
direction of the incoming proton. Then the rapidities in the laboratory frame
and in the center of mass frame are connected by the usual formula
\beq
y_{lab} = y_{cm} + {1\over 2}\log{{E_p}\over{E_\gamma}}
\eeq
We use the MRSA\scite{mrsa} structure function set for the
proton and the ACFGP-mc\scite{acfgp} one for the photon.
The $\Lambda_5^\msbar$ is fixed at the central value
in the MRSA set, namely 151 MeV and the mass of the charm quark is
assumed 1.5 GeV.

Then fig.~\ref{fig1} shows how the fixed order calculation (from here on
referred to as FMNR) and the PFF one do distribute differently the overall
cross
section between the direct and the resolved contributions. The bidifferential
cross section $\rd\sigma/\rd\pt^2\rd y$ is considered at the
rapidity\footnote{The PFF approach actually deals with the pseudorapidity
$\eta$. In the large $\pt$ limit however, where $\pt\gg m$, the two quantities
are coincident.} value $y_{lab} = 1$. In these plots the
r/f scale is taken equal to the reference value
defined in (\ref{muref}). Namely, $\mu = \xi\mu_{ref}$ with $\xi = 1$.

By looking at the plots we immediately appreciate that the direct and resolved
contributions separately do behave differently in the two approaches. The FMNR
direct part tend to be higher than the PFF one. The opposite happens in the
resolved part, the PFF result being markedly higher, especially in the
large $\pt$ region. The total cross section however, where the two components
have been
added together, displays a good similarity between the FMNR and the PFF
results.
The two curves are practically coincident up to $\pt$ values around 10-15 GeV,
and above these values the PFF one is only slightly lower. This deviation is
due
to the multiple gluon emission from the final state charm, which becomes
relevant for $\pt\gg m$. This effect is of course not described by the fixed
order NLO calculation, which only deals with one gluon emission and therefore
predicts an overall harder charm fragmentation function.

The different descriptions that FMNR and PFF give of the direct and resolved
part separately  are due to the different treatment of the
diagrams
in fig.~\ref{phosplit}, where the photon splits into a $c\bar c$ pair.
In the FMNR
calculation the diagram a) is entirely included in the NLO direct part, the
finite mass of the charm quark preventing the splitting from being singular.
Consistently the diagram b), i.e a scattering
initiated by the heavy quark,
does not appear in the resolved part (we would have a double counting
otherwise). In
the PFF approach on the other hand the charm is massless, and the process of
photon splitting has to be separated into two. Above the factorization scale
$\mug$
the contribution is assigned to the direct component, while
below this scale the
splitting process is instead considered non perturbative: it is taken into
account via the photon structure function $F_\gamma^c$ (fig.~\ref{phosplit}b)
and assigned therefore
to the resolved
part. This different separation explains why in the PFF approach
the direct component is
smaller and the resolved one bigger than in the FMNR one.

The direct and resolved components of the cross section
are shown in fig.~\ref{fig1_ydist}, where the rapidity
distribution at $\pt = 10$ and 20 GeV is plotted. Once again, the PFF direct
term falls below the FMNR one, due to the subtraction of part of the photon
splitting to $c\bar c$. By considering the resolved
part, on the other hand, we see that FMNR
displays the typical symmetrical bell shape usually observed in
hadroproduction.
The PFF shows instead a peak in the low rapidity region, similarly to what
happens in the direct component. Still, after adding the two contributions
the two approaches show a remarkable agreement, except at the very
border of phase space where PFF generally produces a lower cross section. This
discrepancy is due to the fragmentation functions being here probed exclusively
at values of $z$ near one. In this region they display an unphysical
singularity, and resummation of Sudakov terms
is needed to ensure a proper behaviour\scite{melenason}.

Next we consider the scale sensitivity of the two contributions in the two
approaches.
To this aim we recall that in the case of hadroproduction the PFF approach
leads to a reduced sensitivity to the r/f scales,
improving considerably the theoretical uncertainty at large $\pt$.
Fig.~\ref{fig2_direct} shows the uncertainty band given by
varying the f/r scale
between $0.25\;\mu_{ref}$ and $2\;\mu_{ref}$ in the direct component. In both
approaches $\xi = 2$ gives the lower curve.This is  coherent with the
definition  of ``direct contribution'' as what is above the factorization
scale.
Then raising
this scale, less room is left of course for this process to take place.
{}From the same figure we also notice a much larger
sensitivity in the PFF approach. This is due to the fact
that in this case a change of the factorization scale also affects the charm
content of the
photon. In FMNR, on the other hand, this process is not factorized and
therefore
its large contribution to the cross section is not affected by a variation of
$\mug$.

Fig.~\ref{fig2_resolved} shows the analogous band for the resolved component.
For FMNR the $\xi
= 2$ choice still gives the lower curve. This is consistent with what was
observed when studing the hadroproduction case\scite{cg}. For PFF, on the
contrary, $\xi = 2$ now gives the higher curve, at least in the large $\pt$
region. This is related to the fact that the photon splitting into $c\bar c$
is assigned to the resolved component below the factorization scale.

Finally, fig.~\ref{fig2_total} shows the scale sensitivity of the
overall cross section.
Due to the opposite behaviours of the two partial contributions previously
considered the two approaches can be seen to display similar sensitivity.

A more detailed analysis of the scale dependence is shown in
fig.~\ref{fig3_scaledep}, where the cross section is
plotted for three different $\pt$ values (10.5, 20.5 and 30.5 GeV)
as a function of the f/r scale in the range $\xi = $0.25--2.
As previously noted,
the dependencies are pretty similar, the PFF approach being slightly less
sensitive. For comparison a PFF result obtained using leading order kernel
cross
sections and $\as$ and fragmentation functions evolutions has been plotted. It
predicts a lower cross section and displays, as expected, a larger scale
sensitivity.

To conclude, we have considered the
open charm production in $\gamma p$ collisions to next-to-leading order
within the Perturbative Fragmentation Functions formalism, to allow
the resummation of $\as\log(\pt^2/m^2)$ terms. Both direct and
resolved components have been considered in detail. In the
large $\pt$ region $(\pt > m)$ the results are found to be consistent with the
fixed order NLO calculations, small discrepancies being found for very large
$\pt$ and at the edge of phase space. The two approaches differ in
the definition and the relative contribution of the direct and resolved terms,
but essentially agree on their sum. The resummation is found to lead to a
slightly reduced sensitivity to the choice of the
renormalization/factorization scale.

\noindent
{\sl Note added: }After this paper was written it came to our attention the
work \cite{kkks}, where a similar analysis was performed.

\vspace{1cm}
{\bf Acknowledgements.} We thank M. Fontannaz and  S. Frixione for having
provided us with their
codes and for the helpful assistance in using them.

\newpage

\begin{figure}[t]
\begin{center}
\parbox{12cm}{
\caption{\label{fig1}\small Comparison between the fixed order
(FMNR) and the Perturbative Fragmentation Function (PFF) prediction
of the charm photoproduction $\pt$ spectrum.}
}
\end{center}
\end{figure}

\begin{figure}[t]
\begin{center}
\parbox{12cm}{
\caption{\label{fig1_ydist}\small Comparison between the FMNR and the PFF
prediction for the rapidity distribution in charm photoproduction. The upper
curves refer to $\pt = 10$~GeV, the lower ones to $\pt = 20$~GeV.}
}
\end{center}
\end{figure}

\begin{figure}[t]
\begin{center}
\parbox{12cm}{
\caption{\label{phosplit}\small The two different ways a charm can come from
a photon: via pointlike coupling (a) or via the photon structure function (b).}
}
\end{center}
\end{figure}

\begin{figure}[t]
\begin{center}
\parbox{12cm}{
\caption{\label{fig2_direct}\small Scale sensitivity of the FMNR and PFF
prediction for the direct contribution to charm photoproduction.}
}
\end{center}
\end{figure}

\begin{figure}[t]
\begin{center}
\parbox{12cm}{
\caption{\label{fig2_resolved}\small Scale sensitivity of the FMNR and PFF
prediction for the resolved contribution to charm photoproduction.}
}
\end{center}
\end{figure}

\begin{figure}[t]
\begin{center}
\parbox{12cm}{
\caption{\label{fig2_total}\small Scale sensitivity of the FMNR and PFF
prediction for charm photoproduction. Both the direct and resolved component at
the NLO are here taken into account.}
}
\end{center}
\end{figure}

\begin{figure}[t]
\begin{center}
\parbox{12cm}{
\caption{\label{fig3_scaledep}\small Scale sensitivity of the charm
photoproduction cross
section as a function of the renormalization/factorization scale. ``PFF LO''
has
been obtained with leading order cross section kernels and $\as$ and
fragmentation functions evolutions.}
}
\end{center}
\end{figure}


\newpage

\begin{thebibliography}{99}
\def    \nuke   #1#2#3{{Nucl. Phys.} {\bf B#1}  (#2) #3}
\def    \pl     #1#2#3{{Phys. Lett.} {\bf #1} (#2) #3}
\def    \plb     #1#2#3{{Phys. Lett.} {\bf B#1} (#2) #3}
\def    \prl    #1#2#3{{ Phys. Rev. Lett.} {\bf #1}  (#2) #3}
\def    \pr     #1#2#3{{Phys. Rev.} {\bf #1}  (#2) #3}
\def    \prd    #1#2#3{{Phys. Rev.} {\bf D#1}  (#2) #3}
\def    \zeit   #1#2#3{{Z. Phys.} {\bf C#1}  (#2) #3}
\def    \cmp    #1#2#3{{Comm. Math. Phys.} {\bf #1}  (#2) #3}
   \bibitem{EN}
     R.K. Ellis and P. Nason, \nuke{312}{1989}{551}
%
  \bibitem{smith}
      J Smith and W.L. van Neerven, \nuke{374}{1992}{36}
%
  \bibitem{NDE}
    P. Nason, S. Dawson and R.K. Ellis, Nucl. Phys. {\bf B303} (1988) 607\\
    P. Nason, S. Dawson and R.K. Ellis, Nucl. Phys. {\bf B327} (1989) 49
%
  \bibitem{wim}
       W. Beenakker, H. Kuijf, W.L. van Neerven and J. Smith,
       Phys. Rev. {\bf D40} (1989) 54\\
       W. Beenakker, W.L. van Neerven, R.~Meng, G.A.~Schuler and J.~Smith,
       Nucl. Phys. {\bf B351} (1991) 507


   \bibitem{mnr}
        M.L. Mangano, P. Nason and G. Ridolfi, \nuke{373}{1992}{295}
%
  \bibitem{fmnr}
           S. Frixione, M.L. Mangano, P. Nason and G. Ridolfi,
           \nuke{412}{1994}{225}
%
     \bibitem{michael}
     M. Kr\"amer, J. Zunft, J. Steegborn and P.M. Zerwas, \plb{348}{1995}{657}
%
    \bibitem{cwz} J. Collins, F. Wilczek and A. Zee, \prd{18}{1978}{242}\\
                  W.J. Marciano, \prd{29}{1984}{580}
%
     \bibitem{olness}
     M.A.G.~Aivazis, F.I.~Olness  and Wu-Ki Tung,
       \prd{50}{1994}{3085}\\
     M.A.G.~Aivazis, J.C.~Collins, F.I.~Olness and Wu-Ki Tung,
       \prd{50}{1994}{3102}\\
     F.I. Olness and S.T. Riemersma, \prd{51}{1995}{4746}
%
     \bibitem{cg}
     M. Cacciari and M. Greco, Nucl. Phys. {\bf B421} (1994) 530
%
  \bibitem{melenason}
    B. Mele and P. Nason,  Nucl. Phys. {\bf B361} (1991) 626
%
   \bibitem{gv}
     M. Greco and A. Vicini, \nuke{415}{1994}{386}
%
  \bibitem{ellis}
       R.K. Ellis, M.A. Furman, H.E. Haber and I. Hinchliffe,
       \nuke{173}{1980}{397}
%
  \bibitem{aversa}
       F. Aversa, P. Chiappetta, M. Greco and J.Ph. Guillet,  Nucl. Phys.
       {\bf B327} (1989) 105
%
  \bibitem{aurenche}
        P. Aurenche, R. Baier, A. Douiri, M. Fontannaz and D. Schiff,
        \nuke{286}{1987}{553}
%
  \bibitem{mrsa}
         A.D. Martin, R.G. Roberts and W.J. Stirling, \prd{50}{1994}{6734}
%
  \bibitem{acfgp}
        P. Aurenche, P. Chiappetta, M. Fontannaz, J.Ph. Guillet and E. Pilon,
         \zeit{56}{1992}{589}
%
\bibitem{kkks}
      B.A. Kniehl, M. Kr\"amer, G. Kramer and M. Spira, DESY~95-098
\end{thebibliography}
\end{document}